\documentclass[a4paper,12pt]{article}
\usepackage[centertags]{amsmath}
\usepackage{amssymb}

\usepackage{graphicx}
\usepackage{epsfig}
\usepackage{ulem}
\usepackage[english]{babel}
\usepackage{array}
\usepackage{amsthm}
\usepackage{latexsym}

\usepackage{yfonts}
\usepackage{mathrsfs}
\usepackage[mathcal]{euscript}

\pdfoutput=1
\usepackage{epsfig}
\usepackage{jheppubm}
\usepackage{mathdots}
\usepackage{MnSymbol}
\usepackage{multirow}

\definecolor{darkraspberry}{rgb}{0.53, 0.15, 0.34}

\definecolor{darkblue}{rgb}{0., 0, 1}

\definecolor{dgreen}{rgb}{0.,0.6,0.}

\definecolor{aquamarine}{rgb}{0.8,0.0,0.8}

\newcommand\smallO{
  \mathchoice
    {{\scriptstyle\mathcal{O}}}
    {{\scriptstyle\mathcal{O}}}
    {{\scriptscriptstyle\mathcal{O}}}
    {\scalebox{.7}{$\scriptscriptstyle\mathcal{O}$}}
  }

\newcommand{\be}{\begin{equation}}
\newcommand{\ee}{\end{equation}}
\newcommand{\bea}{\begin{eqnarray}}
\newcommand{\eea}{\end{eqnarray}}

\title{No Violation of Bell-CHSH Inequalities at Large Distances}
\author{Timofei Rusalev, Daniil Stepanenko and  Igor Volovich}

\affiliation{Steklov Mathematical Institute, Russian Academy of Sciences,\\Gubkina str. 8, 119991, Moscow, Russia}

\emailAdd{rusalev@mi-ras.ru, dstepanenko@mi-ras.ru, volovich@mi-ras.ru}

\abstract{The usual derivation of the violation of Bell-type inequalities can be applied actually only for small distances between detectors. It does not take into account the dependence of the quantum mechanical wave function on space-time variables. We study the behavior of entangled photons obtained in spontaneous parametric down-conversion (SPDC) experiments and show that at large distances there is in fact no violation of the Bell-CHSH inequalities. We show that the initial entangled states become disentangled at large space-like distances. This does not contradict the violation of Bell inequalities observed at small distances between detectors.
We propose an experiment to study the dependence of the quantum correlation function and Bell value on increasing distance between detectors. We predict that these quantities decrease inversely proportional to the increase of the distance between the detectors.}

\begin{document}
\maketitle
\section{Introduction}
The paradox of Einstein, Podolsky and Rosen (EPR) was put forward as an argument in favor of the fact that quantum mechanics
could not be a complete theory but had to be supplemented by additional variables. These additional variables were to restore causality and locality to the theory \cite{Einstein:1935rr}. The further discussion of EPR paradox was made by many authors and in particular by Bell \cite{Bell:1964kc}.

The Bell type of quantum non-locality is an important phenomenon for understanding the foundations of quantum theory. It also plays a significant role in quantum technologies \cite{Aspect1, Aspect2,Clauser:1978ng, Zeilinger}, see for example \cite{ChinaCosmos,EuropaCosmos,Japan,VolOhya, Acc, Trushechkin:2020azr, Teretenkov:2022cdd, Pechen:2002cs,Teretenkov:2019cdd}.

However usually dependence on the space-time variables in this discussion is neglected. In this paper, based on the previous works \cite{Volovich:2000zc,Volovich:2002fr, Baranov:2002db, KNNV}, we reconsider the spatial dependence of quantum mechanical wave-function in Bell-type experiments and show that in fact there is no violation of the the Bell inequalities if the distance between detectors is large enough.

The purpose of this paper is to show that if in spontaneous parametric down-conversion (SPDC) experiments with collinear emission of a photon of type II one takes into account possible effects of entanglement of the spin part with spatial variables in the quantum correlation function of the field operator of quantize electromagnetic field, which is described by the two-photon amplitude, then at sufficiently large distances between detectors it can be represented as expectation of local classical random variables.

We proceed under the assumption that in quantum field theory the general property is that there is a decay of the correlation. As the distance between points corresponding to operator insertions increases, the correlation function tends to zero. See for example \cite{Streater:1989vi}. 
This is not observed at the level of correlations such as Bell inequalities for spin correlation function, which can be written as an  inequality
\begin{equation}\label{eq:Bell-theorem}
    (a \cdot b) \neq \int \xi(a,\lambda)\eta(b,\lambda) d\mu( \lambda),
\end{equation}
where $a = (a_1, a_2, a_3)$ and $b = (b_1, b_2, b_3)$ are two unit vectors in three-dimensional space. This  inequality is the Bell theorem. Namely, it means that the quantum correlation function, i.e. the LHS of ~\eqref{eq:Bell-theorem}, cannot be represented as a classical correlation function on the  RHS 
 of ~\eqref{eq:Bell-theorem} under the assumption that classical processes with classical random variables satisfy $|\xi|\leq 1$, $|\eta| \leq 1$. The random variables $\xi, \eta$ are ordinary functions of the $a, b$ which are second additional variables, and the integral in \eqref{eq:Bell-theorem} is taken over the positive probability measure $\int d\mu( \lambda) = 1$.

If we consider a pair of particles with half spin formed in a singlet spin state and moving freely to two detectors, where we neglect the spatial part of the wave function, we obtain a Hilbert space $C^2 \bigotimes C^2$. The quantum mechanical correlation function of the two spins in the singlet state can be described as
\begin{equation}
\label{CorSpin}
    E_{spin}(a,b)=\langle\psi_{spin}|\sigma \cdot a \bigotimes \sigma \cdot b|\psi_{spin} \rangle =-a \cdot b,
\end{equation} 
where $\sigma = (\sigma_1, \sigma_2, \sigma_3)$ are Pauli matrices. Following the Bell theorem, we know that the correlation function~\eqref{CorSpin} cannot be represented as a classical correlation function as follows 
\begin{equation}
\label{probclas}
P(a,b)=\int \xi(a,\lambda)\eta(b,\lambda) d\mu( \lambda).
\end{equation}

The Clauser-Horne-Shimony-Holt (CHSH) inequality, as purely mathematical and completely independent of physical interpretations, 
claims that the quantity $S_{classical}$,
\begin{equation}
\label{CHSH22}
   S_{classical} = |P(a,b)-P(a,b')+P(a',b)+P(a',b')|, 
\end{equation}
 satisfies $$S_{classical}\leq 2.$$ On the other hand, for the quantum case there is a Bell value
\begin{equation}
    S_{spin} = |E_{spin}(a,b)-E_{spin}(a,b')+E_{spin}(a',b)+E_{spin}(a',b')|.
\end{equation}
Thus, an obvious contradiction arises if we assume that the quantum correlation function without dependence on spatial part can be represent as a classical correlation function. It is well known, that we can choose state $|\psi_{spin} \rangle$ and $a,b$, such that the Bell value is $S_{spin}=2\sqrt{2}$. Therefore, we cannot present $E_{spin}(a,b)$ in the form of $P(a,b)$, see \eqref{probclas}. This fact is often called a violation of the Bell inequalities. 

One of possible reasons for this result is the lack of dependence on the spatial part.

\subsection{Bell inequalities with space-time dependence}

Let us calculate the correlation function using the following scheme. We have 2 detectors, each in its own region $\mathcal{O_A}$ and $\mathcal{O_B}$. To investigate the probability of a spin being directed in a some direction, we need to write a projectors $P_{\mathcal{O_A}}$ and $P_{\mathcal{O_B}}$ on the regions $\mathcal{O_A}$ and $\mathcal{O_B}$.
That is, the spin-space dependent correlation function that includes a projector onto the regions $\mathcal{O_A}$ and $\mathcal{O_B}$ has the following form
\begin{equation}
\label{E-spin-space-}
    E^{spin}_{space}(a,\mathcal{O_A},b,\mathcal{O_B})= \langle \psi|\sigma \cdot a P_{\mathcal{O_A}} \otimes \sigma \cdot b P_{\mathcal{O_B}}|\psi \rangle.
\end{equation}
Spin state prepared in special Bell-type space $\mathbb{C}^2 \bigotimes \mathbb{L}^2( \mathbb{R}^3) \bigotimes\mathbb{C}^2 \bigotimes \mathbb{L}^2(\mathbb{R}^3)$, note also that $ ||\sigma \cdot a P_{\mathcal{O_A}}|| \leq 1$ and $||\sigma \cdot b P_{\mathcal{O_B}}|| \leq 1$. Therefore we consider the corresponding classical correlation function 
\begin{equation}
   \label{probspaceclas}
P(a,\mathcal{O_A},b,\mathcal{O_B})=\int \xi(a,\mathcal{O_A},\lambda)\eta(b,\mathcal{O_B},\lambda) d\mu( \lambda),
\end{equation} 
where $| \xi(a,\mathcal{O_A},\lambda)| \leq 1$ and $|\eta(b,\mathcal{O_B},\lambda)| \leq 1$. We define the classical Bell value in the following way
\begin{equation}
\begin{aligned}
S^{classical}_{space}&=|P(a,\mathcal{O_A},b,\mathcal{O_B})-P(a,\mathcal{O_A},b',\mathcal{O_B'})\\
&+P(a',\mathcal{O_A'},b,\mathcal{O_B})+P(a',\mathcal{O_A'},b',\mathcal{O_B'})|.
\end{aligned}
\end{equation}
One has the modification of the Bell-CHSH inequality $ S^{classical}_{space} \leq 2$.
In the following representation, one can see that the quantum correlation function is just the average value of the Hermitian operator of the observable variable $Q=\sigma \cdot a P_{\mathcal{O_A}} \otimes \sigma \cdot b P_{\mathcal{O_B}}$.

Now we introduce an analog of the classical Bell value for the spin-space correlation function, which is 
\begin{equation}
\begin{aligned}\label{eq:s-spin-space}
    &S^{spin}_{space}=|E(a,\mathcal{O_A},b,\mathcal{O_B})-E(a,\mathcal{O_A},b',\mathcal{O_B'})\\
&+E(a',\mathcal{O_A'},b,\mathcal{O_B})+E(a',\mathcal{O_A'},b',\mathcal{O_B'})|.
\end{aligned}
\end{equation}
Now we would like to emphasize, that in this case we cannot conclude that there is a contradiction between the classical and quantum Bell values. It is because for given state $|\psi \rangle$ if distance between detectors will increase, then $E^{spin}_{space} \to 0$. If one takes the quantum correlation function and inserts the partition of unity we get the following representation
\begin{equation}
    E^{spin}_{space}(a,\mathcal{O_A},b,\mathcal{O_B})= \int \xi(a,\mathcal{O_A},\lambda)\eta(b,\mathcal{O_B},\lambda) d\mu( \lambda).
\end{equation}
In particular, it means that the Bell-CHSH inequalities are valid at large distances, so $S^{spin}_{space}\leq 2$ takes place. 
We believe it is important to conduct the experiment over long distances and detect the disentanglement effect over long distances. For example, we can take a fixed source emitting photons and vary the distance between detectors and study the correlation.
We predict that the Bell value will decrease with increasing distance between detectors.
\subsection{Example}
One theoretical example of disentanglement can be found in the following case. Consider the quantum correlation function that describes localized measurements of spins in detectors located in the $\mathcal{O_A}$ and $\mathcal{O_B}$, with the assumption that the wave-function becomes factorized in the following way $\psi= \psi_{spin} \phi (r_{\mathcal{A}},r_{\mathcal{B}})$
\begin{equation}
 E^{spin}_{space}(a,\mathcal{O_A},b,\mathcal{O_B})=g(\mathcal{O_A},\mathcal{O_B})E_{spin}(a,b),
\end{equation}
where the function $g$ can be written as
\begin{equation}
\label{gint}
    g(\mathcal{O_A},\mathcal{O_B})=\int |\phi (r_{\mathcal{A}},r_{\mathcal{B}})|^2 dr_{\mathcal{A}}dr_{\mathcal{B}},
\end{equation}
and describes the correlation of particles in space. It is the probability to find one particle in the region $\mathcal{O_A}$ and another particle in the region $\mathcal{O_B}$. So from here we see the natural occurrence of the limiting function $|g|\leq 1$. 

Here the spatial part is taken into account, and in the simple case it gives $g$, which tends to zero as the distance between the regions $\mathcal{O_A}$ and $\mathcal{O_B}$ increases \cite{Volovich:2002fr}. This means that the quantum correlation function with dependence on the spatial and spin part can be represented as a classical one. So, the Bell inequalities are always satisfied and they are not violated.

Therefore, as it was established in \cite{Volovich:2002fr}, if the distance between detectors is large, it is possible to represent quantum space dependent correlation function as a classical correlation and therefore from the Bell-type reasoning instantaneous long-range action does not arise here. As for the instantaneous information transfer we need to take into account not only space but also time dependence. This is the quantum non-locality problem.

One should remember that in quantum field theory there is a condition of micro-locality
\begin{equation}
    [\phi_i(x,t), \phi_j(y,\tau)]=0,
\end{equation}
where the space-time insertion points of the fields are space-like separated, which implies that information cannot propagate faster than the speed of light. This contradicts the results of the Bell inequality most likely because of ignorance of the spatial part.
In what follows, the field operator will be the quantized electromagnetic field.

\section{No violation of Bell-CHSH inequalities for the photons SPDC}
Spontaneous parametric down-conversion is a fundamental process in quantum optics, that has practical applications in the controlled generation of entangled states of photons \cite{Zubary}. Understanding the influence of pump coherence on the entanglement produced in SPDC is crucial for both theoretical research and practical implementation. In the study conducted in \cite{Li:2022azd} the focus was on  long-distance quantum communication schemes in SPDC.

SPDC usually requires a coherent laser beam as a pump source, but recently a breakthrough in quantum entanglement \cite{Zhang:2022bsf} has been achieved by successfully generating polarization-entangled photon pairs using light-emitting diodes (LEDs) as a pump source.

\subsection{Description of the experiment}
The LED source has low spatial coherence, but the type-II phase-matching condition in the nonlinear crystal used in the experiment filtered specific frequencies and wavelengths of the LED light to participate in the down-conversion process, allowing for the generation of localized polarization Bell states.

The degree of polarization entanglement induced by the LED light was characterized using the violation of Bell inequality. The  Bell value of $S = 2.33 \pm 0.097$ has been achieved by experiment \cite{Zhang:2022bsf}, surpassing the classical bound of $S = 2$ and demonstrating quantum entanglement. This experiment opens up possibilities for long-distance  quantum communications in outer space.

\subsection{Setup}
We assume the presence of an ideally calibrated detector characterized by minimal error and this ideal conditions do not depended on the distance  between detectors $D_s$ and $D_i$ located at $\boldsymbol{r}_s$ and $\boldsymbol{r}_i$, where the indices $s$ and $i$ correspond to the signal and idler photons appearing in SPDC, respectively. Consequently, quantum efficiencies can be set $\alpha = 1$, where $\boldsymbol{r}_j$ is defined as $\boldsymbol{r}_j\equiv (\boldsymbol{\rho}_j,z )$, and $j=i, s$. We consider the positive and negative frequency parts of the quantized electromagnetic fields $\hat E^{(+)}(\theta,\boldsymbol{r})$ and $\hat E^{(-)} (\theta,\boldsymbol{r})$ \cite{Zubary,Glaber}.

In what follows, $H$ and $V$ denote the polarization states of the photon, and $\left|H_s,\boldsymbol{q}_s\right>$ and $\left|V_i,\boldsymbol{q}_i\right>$ represent the states for the mode with polarization and transverse wave vector $\boldsymbol{q}_s$, $\boldsymbol{q}_i$ of the signal and idler. 
Space-spin dependent quantum correlation function $R$ can be described by the following representation 
\begin{equation}
R_{si}(\theta_s,\theta _i,\boldsymbol{r}_s,\boldsymbol{r}_i)=Tr [\rho\hat{E}_{s}^{(-)}(\theta _s,\boldsymbol{r}_s)\hat{E}_{i}^{(-)}(\theta _i,\boldsymbol{r}_i) \hat{E}_{i}^{(+)}(\theta_i,\boldsymbol{r}_i) \hat{E}_{s}^{( + )}(\theta_s,\boldsymbol{r}_s)].
\end{equation}
The two-photon entangled state in this experiment has the form
\begin{equation}
\label{density matrix}
\begin{aligned}
&\hat{\rho}=|A|^2\int{d\boldsymbol{q}_sd\boldsymbol{q}_id\boldsymbol{q}_{s}^{'}d\boldsymbol{q}_{i}^{'}}\langle G( \boldsymbol{q}_s+\boldsymbol{q}_i) G^*( \boldsymbol{q}_{s}^{'}+\boldsymbol{q}_{i}^{'})\rangle \\
&\times\operatorname{sinc}^2( \frac{\varDelta qL}{2}) \frac{1}{\sqrt{2}}(|H_s,\boldsymbol{q}_s\rangle|V_i,\boldsymbol{q}_i\rangle +|V_s,\boldsymbol{q}_s\rangle|H_i,\boldsymbol{q}_i\rangle)
\\
&\times \frac{1}{\sqrt{2}}({\langle H_{s'},\boldsymbol{q}_{s}^{'}|\langle V_{i'},\boldsymbol{q}_{i}^{'}|+\langle V_{s'},\boldsymbol{q}_{s}^{'}|\langle H_{i'},\boldsymbol{q}_{i}^{'}|}).
\end{aligned}
\end{equation}
Additional degrees of freedom, such as the spatial freedom of the photon in this mode, are encoded in its transverse wave vector. $|A|^2$ remains a constant contingent on physical constants. The term $\langle G(\cdot)G^*(\cdot)\rangle$ represents the angular correlation function of the pump field, $\boldsymbol{q}_s$ and $\boldsymbol{q}_i$ denote the transverse wave vectors of the signal and idler fields. The function $\operatorname{sinc}^2(\varDelta qL/2)$ signifies the phase matching condition for the nonlinear crystal.

According to the definition of the density matrix \eqref{density matrix}, the state is determined by the mode function, which is inextricably linked to the pump angular spectrum, the crystal phase matching function, and the polarization projection. If the signal photon is detected at position $\boldsymbol{r}_s$ with a polarization angle $\theta_s$ and the idler photon is detected at position $\boldsymbol{r}_i$ with a polarization angle $\theta_i$, then the count rate of the two-photon state coincidence can represent the space-spin dependent quantum correlation function \cite{Zhang:2022bsf}
\begin{equation}
\label{Correlator}
\begin{aligned}
&R_{si}(\theta _s,\theta _i ,\boldsymbol{r}_s,\boldsymbol{r}_i) =\Bigg|| A|^2e^{ik_p(\boldsymbol{r}_s+\boldsymbol{r}_i )}\Bigg[\frac{A_p\pi \ell_c{\sigma _0}^2}{z\sqrt{4{\sigma _0}^2+\ell _{c}^{2}}}\Bigg]^2\\
& \times\Bigg[C_1C_2\cdot \frac{\pi z^2}{2k_{p}^{2}\sigma _z\delta}\sin 2\theta _s\sin 2\theta _i\\
&+\sqrt{\frac{2\pi}{k_{p}^{2}\delta ^2}}(C_{1}^{2}\cos ^2\theta _s\sin ^2\theta_i+C_{2}^{2}\sin ^2\theta _s\cos ^2\theta_i) \Bigg]\Bigg|,
\end{aligned}
\end{equation}

In this context, $k_p$ represents the pump wave vector, $\sigma _0$ denotes the beam size, and $\sigma _{z}$ is the beam size at position $z$ along the propagating direction. The relationship between these quantities is given by $\sigma _z={{z\sqrt{\ell _{c}^{2}+4\sigma _{0}^{2}}}/{2k_p\sigma _0\ell _c}}$, where $ \ell_c $ stands for the transverse coherence length. The effective spectral width is defined as $\delta =2\ell _c\sigma _z/\sqrt{4{\sigma _z}^2+\ell _{c}^{2}}$ . Additionally, $C_1$ and $C_2$ represent the detection efficiency of the entire system, and $A_p$ is a constant induced by the cross-spectral density function of a Gaussian Schell Model (GSM) pump beam.

\begin{figure}[t!]\centering
    \includegraphics[width=0.675\textwidth]{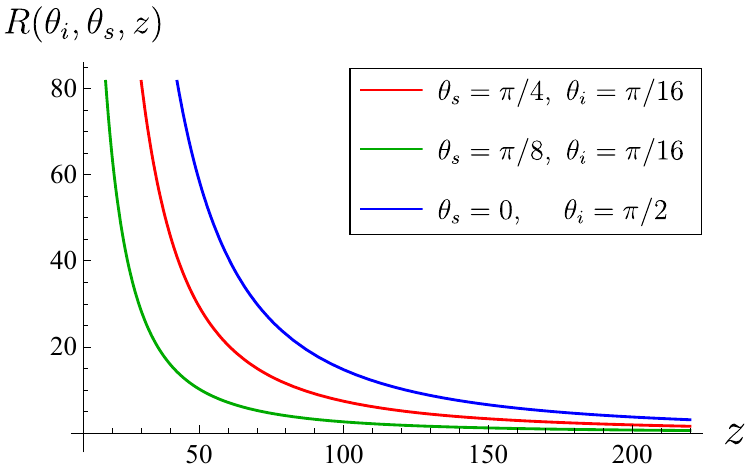}
    \caption{The qualitative behavior of the decreasing quantum correlation function \eqref{Correlator} with increasing distance between detectors.}
 \label{fig:decreasing-of-correlator}
\end{figure}

From \eqref{Correlator} it becomes evident that the count rate does not explicitly depend on $\boldsymbol{r}_i$, $\boldsymbol{r}_s$, only up to a phase factor. However, both the beam size $\sigma_z$ and the spatial coherence $\delta$ exert an influence on the count rate, that depend on the distance along the direction of propagation $z$. From \eqref{Correlator} one can see that the quantum correlation function equation on asymptotic became a proportional to at $z\to \infty$ is
\begin{equation}
\label{null}
    R(\theta _s,\theta _i ,\boldsymbol{r}_s,\boldsymbol{r}_i) = \frac{\mathcal{K}}{z}+\smallO (1/z),
\end{equation}
where constant $\mathcal{K}$ depends on angles of polarization and initial beam size. The qualitative behavior of the decreasing quantum correlation function \eqref{Correlator} with increasing~$z$ is presented in Fig.\ref{fig:decreasing-of-correlator}. The Bell value is expressed through the quantum correlation function \eqref{Correlator} as follows
\begin{equation}
\begin{aligned}
   S_{space}& = |R(\theta _s,\theta _i ,\boldsymbol{r}_s,\boldsymbol{r}_i)-R(\theta _s,\theta _i',\boldsymbol{r}_s,\boldsymbol{r}_i') \\ &+ R(\theta _s',\theta _i ,\boldsymbol{r}_s',\boldsymbol{r}_i) + R(\theta _s',\theta _i',\boldsymbol{r}_s',\boldsymbol{r}_i')|.   
\end{aligned}
\end{equation}
We see that $S_{space}\to 0$ because at $z\to \infty$  every term $R$, which is included in \eqref{null}, tends to zero.
Therefore there is no violation of the Bell inequalities at large enough distances, so it tends to zero with the distance tends to infinity.\footnote{Note that $S_{space}$ is dimensionful in contrast to $S^{spin}_{space}$, see \eqref{eq:s-spin-space}. We are grateful to A.E. Teretenkov for this remark.}
And we can represent quantum correlation function $R$ as classical correlation function at large distances
\begin{equation}
    R(\theta _s,\theta _i ,\boldsymbol{r}_s,\boldsymbol{r}_i) = \int \xi(\theta _s,r _s,\lambda) \eta(\theta _i,r _i, \lambda) d\mu(\lambda).
\end{equation}
Then we do not have a contradiction between classical and quantum correlation function.
For a similar proof, one can see \cite{Khrennikov:2002wz}. 

It is important to conduct experiments at different distances and try to experimentally find a correlation between detector distances and the Bell value. We expect the Bell value to depend on the distance between detectors. One can check this using the following scheme.  Let us take a measurement at a certain distance and obtain the Bell value. After this, let us  increase the distance between the detectors and repeat the measurement procedure, calculate the Bell value for this distance, and so on. In this way, one can obtain the functional dependence of the Bell value on the distance between the detectors. 
We predict that the Bell value will decrease with increasing distance between detectors. 

\section{Discussions}

We would like to discuss the relationship between our proposal in this paper and the experimental setting. In \cite{Aspect3} for the correlation function of two spins the following formula is considered
\begin{equation}\label{eq:experimental-setting}
    E(a,b)=\frac{N_{++}(a,b) - N_{+-}(a,b) - N_{-+}(a,b) + N_{--}(a,b)}{N_{++}(a,b) + N_{+-}(a,b) + N_{-+}(a,b) + N_{--}(a,b)}.
\end{equation}
From this expression it was proven that the Bell-CHSH inequalities are violated. We interpret this result as concerning the violation of the Bell inequalities only at small distances. Under the assumption that we can create entanglement photon pairs and measure the 4 coincidence frequencies $N_{\pm}(a,b)$ using detectors, we can obtain the polarization correlation coefficient in orientations $a$ and $b$.
This expression corresponds to the expectation value $E_{spin} (a,b)$, see \eqref{CorSpin}.
For a classical Bell value we have the Bell-CHSH inequality $S_{classical} \leq 2$.

In this work, we considered the spin correlation function as a function of the distance between detectors $E^{spin}_{space}(a,\mathcal{O_A},b,\mathcal{O_B})$, see \eqref{E-spin-space-}. For the corresponding classical correlation function $P(a,\mathcal{O_A},b,\mathcal{O_B})$ we have obtained the generalized Bell-CHSH inequality $S^{classical}_{space} \leq 2$. Moreover, we have shown that in the quantum case Bell value $S^{spin}_{space}$ tends to zero when the distance between detectors increases. Therefore, there is no violation of Bell-CHSH inequalities at large distances. Note that we do not involve the rational expression of the correlation function in our consideration.

Furthermore, in \cite{Zhang:2022bsf} the following expression for experimental data is considered
\begin{equation}
\begin{aligned}
E( \theta _s,\theta _i ) =\frac{R( \theta _s,\theta _i ) +R( \theta _{s}^{_{\bot}},\theta _{i}^{_{\bot}} ) -R( \theta _{s}^{_{\bot}},\theta _i ) -R( \theta _s,\theta _{i}^{_{\bot}} )}{R( \theta _s,\theta _i ) +R( \theta _{s}^{_{\bot}},\theta _{i}^{_{\bot}} ) +R( \theta _{s}^{_{\bot}},\theta _i ) +R( \theta _s,\theta _{i}^{_{\bot}} )}.
\end{aligned}
\end{equation}
and the violation the Bell-CHSH inequalities is obtained. We do not consider rational expressions of correlation functions because we cannot establish Bell-CHSH type inequalities for the corresponding classical expressions. Instead, we have shown that the quantum correlation function $R(\theta _s,\theta _i ,\boldsymbol{r}_s,\boldsymbol{r}_i)$ tends to zero when the distance between detectors tends to infinity, $z \to \infty$. In this sense, there is no violation of the Bell-CHSH inequalities.

\section{Conclusions}
The Bell quantum non-locality is investigated in many works. This phenomenon is very important in quantum technology, especially in cryptography.  However, dependence on the space-time variables in this discussion is usually neglected. In this paper, based on the previous works, we reconsider the spatial dependence of the quantum mechanical wave-function in the Bell type experiments and show that in fact there is no violation of the Bell inequalities if the distance between detectors is large enough. We consider the Bell type experiments with entangled spin variables and with entangled photons.
In both cases we find that the correlation function and the Bell value $S^{spin}_{space}$ decrease when the distance between detectors increases. We proposed to conduct experiment for studying the spatial dependence of the Bell value and corresponding correlation function. These values are predicted to decrease with increasing distances between detectors.

\section*{Acknowledgement}
We would like to thank I.Ya.Aref'eva and A.N.Pechen for interest and support of our project, as well as A.S.Trushechkin, A.E.Teretenkov, R.Singh for very helpful and exciting discussions at Quantum Mathematical Physics Seminar at Steklov Mathematical Institute. We also thank Wuhong Zhang for useful communication.



\newpage

\begin{thebibliography}{99}

\bibitem{Einstein:1935rr}
A.~Einstein, B.~Podolsky and N.~Rosen,
``Can quantum mechanical description of physical reality be considered complete?,''
Phys. Rev. \textbf{47} (1935), 777-780
doi:10.1103/PhysRev.47.777  

\bibitem{Bell:1964kc}
J.~S.~Bell,
``On the Einstein-Podolsky-Rosen paradox,''
Physics Physique Fizika \textbf{1}, 195-200 (1964)
doi:10.1103/PhysicsPhysiqueFizika.1.195


\bibitem{Aspect1}
A. Aspect, P. Grangier and G. Roger, Experimental Tests of Realistic Local Theories via Bell Theorem, Phys. Rev. Lett. 47, 460 (1981)

\bibitem{Aspect2}
A. Aspect, Proposed experiment to test the nonseparability of quantum mechanics, Phys. Rev. D 14, 1944 (1976)

\bibitem{Clauser:1978ng}
J.~F.~Clauser and A.~Shimony,
``Bell theorem: Experimental tests and implications,''
Rept. Prog. Phys. \textbf{41}, 1881-1927 (1978)
doi:10.1088/0034-4885/41/12/002

\bibitem{Zeilinger}
 M. Zukowski, A. Zeilinger, M.A. Horne and A.K. Ekert, ‘‘Event-ready-detectors’’ Bell experiment via entanglement swapping, Phys. Rev. Lett. 71, 4287 (1993)


\bibitem{ChinaCosmos}
Yin, J., Li, YH., Liao, SK. et al. Entanglement-based secure quantum cryptography over 1,120 kilometres. Nature 582, 501–505 (2020). https://doi.org/10.1038/s41586-020-2401-y

\bibitem{EuropaCosmos}
de Forges de Parny, L., Alibart, O., Debaud, J. et al. Satellite-based quantum information networks: use cases, architecture, and roadmap. Commun Phys 6, 12 (2023). https://doi.org/10.1038/s42005-022-01123-7

\bibitem{Japan}
Kamimaki, A., Wakamatsu, K., Mikata, K. et al. Deterministic Bell state measurement with a single quantum memory. npj Quantum Inf 9, 101 (2023). https://doi.org/10.1038/s41534-023-00771-z


\bibitem{VolOhya}
Masanori Ohya, Igor Volovich, Mathematical Foundations of Quantum Information and Computation and Its Applications to Nano- and Bio-systems, 
https://doi.org/10.1007/978-94-007-0171-7

\bibitem{Acc}
Luigi Accardi, Yun Gang Lu, Igor Volovich, Quantum Theory and Its Stochastic Limit, https://doi.org/10.1007/978-3-662-04929-7

\bibitem{Trushechkin:2020azr}
A.~Trushechkin, ``Security of quantum key distribution with detection-efficiency mismatch in the multiphoton case,''
Quantum \textbf{6} (2022), 771
doi:10.22331/q-2022-07-22-771
[arXiv:2004.07809 [quant-ph]].

\bibitem{Teretenkov:2022cdd}
Linowski, T., Teretenkov, A. and Rudnicki, Ł., 2022. Dissipative evolution of quantum Gaussian states. Physical Review A, 106(5), p.052206. https://doi.org/10.1103/PhysRevA.106.052206

\bibitem{Pechen:2002cs}
Pechen, Alexander, and Herschel Rabitz. "Teaching the environment to control quantum systems." Physical Review A 73.6 (2006): 062102.

\bibitem{Teretenkov:2019cdd}
A.~E.~Teretenkov,
``Irreversible quantum evolution with quadratic generator: Review,''
doi:10.1142/S0219025719300019
[arXiv:1912.13083 [quant-ph]].

\bibitem{Volovich:2000zc}
I.~V.~Volovich,
``Bell theorem and locality in space,''
[arXiv:quant-ph/0012010 [quant-ph]].

\bibitem{Volovich:2002fr}
I. Volovich, “Quantum cryptography in space and Bell theorem”, Foundations of probability and physics, QP–PQ: Quantum Probab. White Noise Anal., 13, World Sci. Publ., River Edge, NJ, 2001, 364–372, [arXiv:quant-ph/0207050 [quant-ph].

\bibitem{Baranov:2002db}
A. A. Baranov, A. N. Pechen, I. V. Volovich, “Spatial dependence of entangled states and some EPR experiments”, Quantum theory: reconsideration of foundations—2, Mathematical Modelling in Physics, Engineering and Cognitive Science, Vol. 10, ed. A. Yu. Khrennikov, Vaxjo University, 2004, 83--91,
[arXiv:quant-ph/0203152 [quant-ph]].

\bibitem{KNNV}
A. Khrennikov, B. Nilsson, S. Nordebo, I. V. Volovich; Distance dependence of entangled photons in waveguides. AIP Conf. Proc. 29 March 2012; 1424 (1): 262–269. https://doi.org/10.1063/1.3688979

\bibitem{Streater:1989vi}
R.~F.~Streater and A.~S.~Wightman,
``PCT, spin and statistics, and all that,''

\bibitem{Gu2023}
Gu, B., Sun, S., Chen, F., Mukamel, S. (2023). Photoelectron spectroscopy with entangled photons; enhanced spectrotemporal resolution. Proceedings of the National Academy of Sciences of the United States of America, 120(21), e2300541120. https://doi.org/10.1073/pnas.2300541120

\bibitem{Zhang:2022bsf}
W.~Zhang, D.~Xu and L.~Chen,
Phys. Rev. Applied \textbf{19} (2023) no.5, 054079
doi:10.1103/PhysRevApplied.19.054079
[arXiv:2211.00841 [quant-ph]].

\bibitem{Zubary}

Scully, Marlan O., and M. Suhail Zubairy. "Quantum optics." (1999): 648-648.

\bibitem{Li:2022azd}
C.~Li, B.~Braverman, G.~Kulkarni and R.~W.~Boyd,
Phys. Rev. A \textbf{107} (2023) no.4, L041701
doi:10.1103/PhysRevA.107.L041701
[arXiv:2210.16229 [quant-ph]].

\bibitem{Glaber}
The Quantum Theory of Optical Coherence
Roy J. Glauber
Phys. Rev. 130, 2529 – Published 15 June 1963

\bibitem{Khrennikov:2002wz}
A.~Khrennikov and I.~Volovich,
[arXiv:quant-ph/0211078 [quant-ph]].

\bibitem{Aspect3}
Aspect, A. (2002). Bell’s Theorem: The Naive View of an Experimentalist. In: Quantum [Un]speakables. Springer, Berlin, Heidelberg. https://doi.org/10.1007/978-3-662-05032-3\_9

\end{thebibliography}
\end{document}